\documentclass{aa}  

\usepackage{graphicx}
\usepackage{txfonts}
\usepackage[colorlinks=true, allcolors=blue]{hyperref}
\usepackage{orcidlink}
\usepackage[export]{adjustbox}
\usepackage{multicol}
\usepackage{multirow}
\usepackage{enumitem} 
\usepackage{cuted} 
\usepackage{subcaption}
\usepackage{tikz}

\newcommand {\ixpe}{\text{IXPE}\xspace}

\begin{document} 

   \title{Inelastic scattering effects on polarization predictions for cold and warm atmospheres in the X-rays}

   \titlerunning{X-ray polarization from inelastic scattering in atmospheres}

   \author{J.~Podgorn{\'y} \inst{1}\corrauth{jakub.podgorny@asu.cas.cz}\orcidlink{0000-0001-5418-291X},
   L.~Marra\inst{2}\orcidlink{0009-0001-4644-194X},
   M.~Dov{\v{c}}iak \inst{1}\orcidlink{0000-0003-0079-1239},
   R.~Taverna\inst{3}\orcidlink{0000-0002-1768-618X},
   R.~Goosmann\inst{4},
   M.~Gupta\inst{1}\orcidlink{0000-0003-0976-8932},
   F.~Marin\inst{4}\orcidlink{0000-0003-4952-0835},
   G.~Matt\inst{5}\orcidlink{0000-0002-2152-0916},
   J.~Poutanen\inst{6}\orcidlink{orcid=0000-0002-0983-0049},
A.~R\'o\.za\'nska\inst{7}\orcidlink{0000-0002-5275-4096},\and 
   A.~Veledina\inst{6,8}\orcidlink{0000-0002-5767-7253}
   }

   \institute{
   Astronomical Institute of the Czech Academy of Sciences, Bo\v{c}n\'{i} II 1401/1, 14100 Praha 4, Czech Republic
   \and INAF Istituto di Astrofisica e Planetologia Spaziali, Via del Fosso del Cavaliere 100, 00133 Roma, Italy
   \and Dipartimento di Fisica e Astronomia, Universit\`{a} degli Studi di Padova, Via Marzolo 8, 35131 Padova, Italy
   \and Universit\'{e} de Strasbourg, CNRS, Observatoire Astronomique de Strasbourg, UMR 7550, 67000 Strasbourg, France
   \and Dipartimento di Matematica e Fisica, Università degli Studi Roma Tre, Via della Vasca Navale 84, 00146 Roma, Italy
   \and Department of Physics and Astronomy, 20014 University of Turku, Finland
   \and Nicolaus Copernicus Astronomical Center, Polish Academy of Sciences, Bartycka 18, 00-716 Warsaw, Poland
   \and Nordita, KTH Royal Institute of Technology and Stockholm University, Hannes Alfv\'ens v\"ag 12, SE-10691 Stockholm, Sweden
   }

  \authorrunning{J. Podgorn{\'y} et al.}

   \date{Received ...; Accepted ...}

 
\abstract
{X-ray polarization models of accreting compact objects often assume the so-called Chandrasekhar--Sobolev polarization prescription for elastic Thomson scattering in electron-scattering semi-infinite atmospheres. In this work, we examine the effects of inelastic Compton scattering. We assume homogeneous slabs of cold and warm (up to 1 keV) electrons. We use a Monte Carlo approach. 
We discuss the polarization difference arising in the X-rays between the elastic and inelastic scattering assumptions for plane-parallel transmission and reflection, and the role it can play in the interpretation of X-ray polarimetric observations of accreting compact objects.
   }


   \maketitle
%
\nolinenumbers

\section{Introduction}
\label{introduction}

Classical predictions for scattering-induced polarization in plane-parallel atmospheres include 
\begin{itemize}
    \item Chandrasekhar's \citep[Section 68,][]{Chandrasekhar1960} and Sobolev's \citep{Sob63} results for a semi-infinite electron-scattering slab in transmission, often used for emitting optically thick ionized atmospheres,
    \item Chandrasekhar's \citep[Section 70,][]{Chandrasekhar1960} and Sobolev's \citep{Sob63} results for reflection from a semi-infinite electron-scattering slab, often used for reflecting optically thick ionized (diffuse) or neutral (first-order scattering) atmospheres.
\end{itemize}
They presume the Thomson type of elastic scattering.

Many contemporary X-ray models \citep[e.g.,][]{Dovciak2008, Li2009, Schnittman2009, Schnittman2010, Schnittman2016, Zhang2019, Zhang2022, Krawczynski2022, Loktev2022, West2023, Loktev2024} incorporate these results to approximate polarization from reprocessing in accretion-disk atmospheres, which is then used to provide predictions for X-ray polarimeters, such as the \ixpe\ mission operating in 2--8 keV \citep{Weisskopf2022}, or to directly interpret their data. However, it has been noted by several authors \citep[][]{Matt1993, Poutanen1996b, Davis2009, Marra2025, Podgorny2026} that the classical predictions may not hold for reprocessing in accretion-disk atmospheres in the X-rays ($\gtrsim1$ keV) where inelastic Compton scattering effects begin to play a role, similarly to Comptonization in hot coronae \citep[e.g.,][]{SunyaevTitarchuk1985, Poutanen1996}.

In this work, we systematically investigate the departures from the classical predictions due to the Compton recoil in passive cold and warm (up to 1 keV) electron-scattering slabs in the X-ray band. We focus on the role of electron temperature, spectrum of the source, and rest-frame geometry. In Section \ref{compton}, we recapitulate the theory of Compton recoil. In Section \ref{methods}, we provide the methods of our calculations. In Section \ref{results}, we show the results for transmission and reflection. We summarize and discuss the implications in Section \ref{conclusions}.

\section{Theoretical background}\label{compton}

Consider radiation with a Planck spectrum of temperature $T_{\mathrm{BB}}$ propagating
through a slab of thermal electrons with uniform temperature
$T_{\mathrm{e}}$ and Thomson optical depth
$\tau_{\mathrm{es}}$. In the non-relativistic Kompaneets regime,
$kT_{\mathrm{e}}\ll m_{\mathrm{e}}c^2$ and
$h\nu\ll m_{\mathrm{e}}c^2$, the mean fractional energy change of a
photon of energy $E=h\nu$ in one scattering is
\begin{equation}
\left\langle \frac{\Delta E}{E} \right\rangle
\simeq
\frac{4kT_{\mathrm{e}}-E}{m_{\mathrm{e}}c^2},
\label{eq:compton_drift}
\end{equation}
where $k$ is Boltzmann's constant, $h$ is Planck's constant,
$m_{\mathrm{e}}$ is the electron mass, and $c$ is the speed of light \citep{Kompaneets1957,RybickiLightman1979}.
The $4kT_{\mathrm{e}}$ term describes the mean energy gain from the
thermal motion of the electrons, whereas the $E$ term describes
Compton recoil. Photons with $E<4kT_{\mathrm{e}}$ therefore tend to
be up-scattered, while photons with $E>4kT_{\mathrm{e}}$ tend to be
down-scattered. The spectral redistribution produced by thermal Compton scattering is
described, in the non-relativistic Fokker--Planck limit of small
fractional energy changes per scattering, by the Kompaneets equation
\citep{Kompaneets1957,RybickiLightman1979},
\begin{equation}
\frac{\partial n}{\partial t}
=
\frac{n_{\rm e}\sigma_{\rm T}c}{m_{\rm e}c^2}
\frac{1}{E^2}\frac{\partial}{\partial E}
\left\{
E^4
\left[
kT_{\rm e}\frac{\partial n}{\partial E}
+n+n^2
\right]
\right\},
\label{eq:kompaneets}
\end{equation}
where $n(E)$ is the photon occupation
number, $n_{\rm e}$ is the electron number density and 
$\sigma_{\rm T}$ is the Thomson cross-section. The three terms in square brackets describe
thermal diffusion in photon energy, recoil, and stimulated Compton
scattering, respectively
\citep[e.g.,][]{Chluba2008}. For a Planckian radiation field of temperature $T_{\rm BB}$,
\begin{equation}
n_{\rm BB}(E)
=
\frac{1}{\exp(E/kT_{\rm BB})-1},
\label{eq:planck_occupation}
\end{equation}
and therefore
\begin{equation}
kT_{\rm BB}\frac{\partial n_{\rm BB}}{\partial E}
=
-n_{\rm BB}\left(1+n_{\rm BB}\right).
\label{eq:planck_identity}
\end{equation}
Substituting this relation into the Kompaneets operator gives
\begin{equation}
kT_{\rm e}\frac{\partial n_{\rm BB}}{\partial E}
+n_{\rm BB}+n_{\rm BB}^2
=
\left(1-\frac{T_{\rm e}}{T_{\rm BB}}\right)
n_{\rm BB}\left(1+n_{\rm BB}\right).
\label{eq:planck_balance}
\end{equation}
Thus, when $T_{\rm e}=T_{\rm BB}$, the frequency-space Compton flux
vanishes, consistent with the Bose--Einstein equilibrium solution of
the Kompaneets equation
\citep{Kompaneets1957,RybickiLightman1979}. 

In an accretion-disk atmosphere the Kompaneets operator must additionally be coupled to spatial radiative transfer, because the radiation field and electron temperature vary with depth and true absorption and emission are also present. \citet{HubenyEtAl2001} incorporate this physics through a Kompaneets-like Compton source function in their plane-parallel non-LTE disk-atmosphere calculations. For a finite slab, the accumulated redistribution also depends on how
many scatterings occur before escaping. A useful signed estimate is
\begin{equation}
y(E)
\simeq
\frac{4kT_{\mathrm{e}}-E}{m_{\mathrm{e}}c^2}\,
N_{\mathrm{char}},
\label{eq:signed_y}
\end{equation}
where $N_{\mathrm{char}}$ is a characteristic number of scatterings.
For $\tau_{\mathrm{es}}\gg1$, spatial diffusion gives
$N_{\mathrm{char}}\sim\tau_{\mathrm{es}}^2$; the order-of-magnitude
interpolation
$N_{\mathrm{char}}\sim\max(\tau_{\mathrm{es}},\tau_{\mathrm{es}}^2)$
also recovers the optically thin scaling, although the precise value
depends on the photon injection and escape geometry
\citep{RybickiLightman1979,HubenyEtAl2001}. In scattering-dominated atmospheres, photons of different frequencies
thermalize at different depths, so \citet{HubenyEtAl2001} characterize
the formation layer by its electron-scattering optical depth
$\tau_\nu^*$ and local temperature $T_\nu^*$. They define the
positive diagnostic
\begin{equation}
y_\mathrm{eff}
=
\frac{4kT_\nu^*}{m_{\mathrm{e}}c^2}
\max\left[
\tau_\nu^*,(\tau_\nu^*)^2
\right],
\label{eq:davis_y_positive}
\end{equation}
which follows from Equation (\ref{eq:signed_y}) and measures whether repeated Compton scattering is strong enough
to modify the emerging spectrum by means of up- or down-scatterings in the atmospheres transmitting thermal radiation. We aim to demonstrate in a much simpler passive slab set-up that even small values of $y_\mathrm{eff}$, driven by the relative difference between the local electron and radiation temperature and the thickness of the slab, can be associated with a significant modification of polarization of the rest-frame emission in the X-ray band due to the net inelastic scattering effect.

\section{Methods}\label{methods}

\begin{figure} 
\centering
\includegraphics[width=1\linewidth]
{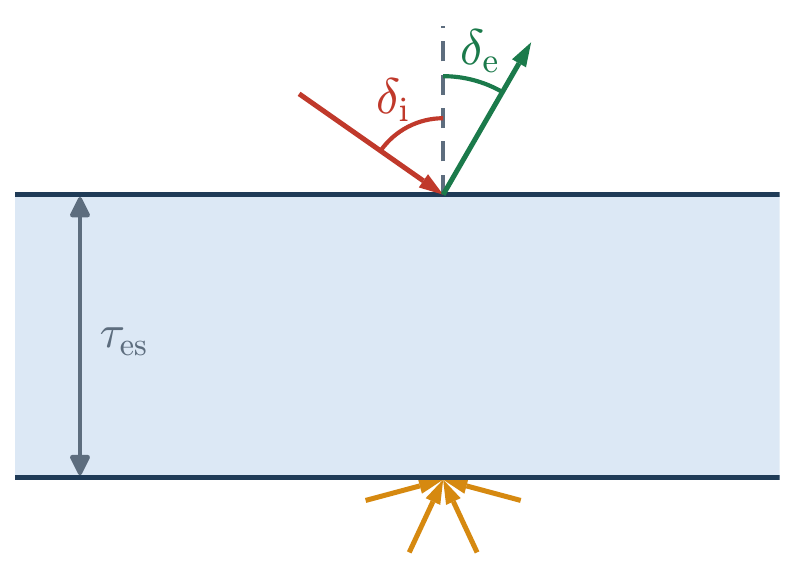}
\caption{Sketch of the plane-parallel geometry assumed.} \label{fig:sketch}
\end{figure}

Figure~\ref{fig:sketch} shows the plane-parallel set up that we adopted to demonstrate the effects of cold and warm Comptonization on X-ray polarization in atmospheres, not only in transmission, but also in reflection. In order to show the main effects, we assume a homogeneous slab of finite $\tau_\mathrm{es}$ and, inside the slab, we assume electrons only; no internal emission or absorption effects or spectral lines are studied. The incident radiation is considered unpolarized and either isotropic from underneath the slab (studied in transmission), or impinging under an inclination angle, $\delta_\mathrm{i}$, measured from the slab normal (studied in reflection). We register the emission inclination angle, $\delta_\mathrm{e}$, measured from the slab normal. The results are averaged in the azimuthal angle. Due to the axial symmetry, Stokes parameter $U$ is zero, and we use the convention of a positive/negative polarization degree (PD), when the electric vector polarization angle is parallel/perpendicular with the slab normal projected to the polarization plane (Stokes parameter $Q$ is positive/negative), respectively.

The results were obtained with a Monte Carlo (MC) method. We used the {\tt STOKES} code \citep{Goosmann_2007, Marin_2012, Marin_2015, Marin_2018_UV}, version 2.36b. The code has been recently updated to include Comptonization \citep[Gupta et al. in prep.; ][]{Podgorny2026}. In the variant that we used, it calculates upon each scattering event a Lorentz boost to the electron's rest frame, a rest-frame Compton scattering, and a transformation back to the laboratory's frame of reference \citep[as described, e.g., in][]{Krawczynski2012}, which was cross-validated to other codes and methods \citep{Poutanen1993, Poutanen1996, Zhang2019}. The electrons are distributed in velocity according to the Maxwell-Jüttner distribution. For the isotropic source in transmission, the photons are angularly distributed according to Equation (4) in \cite{Taverna2021}, i.e., isotropically in specific intensity.


\section{Results}\label{results}

\subsection{Transmission}\label{transmission}

\begin{figure*}[t]
    \centering
    \begin{subfigure}[t]{0.41\textwidth}
        \centering
        \begin{tikzpicture}
            \node[inner sep=0] (img)
                {\includegraphics[width=\linewidth,     
                   trim={0cm 0.35cm 0cm 0cm},
                   clip]{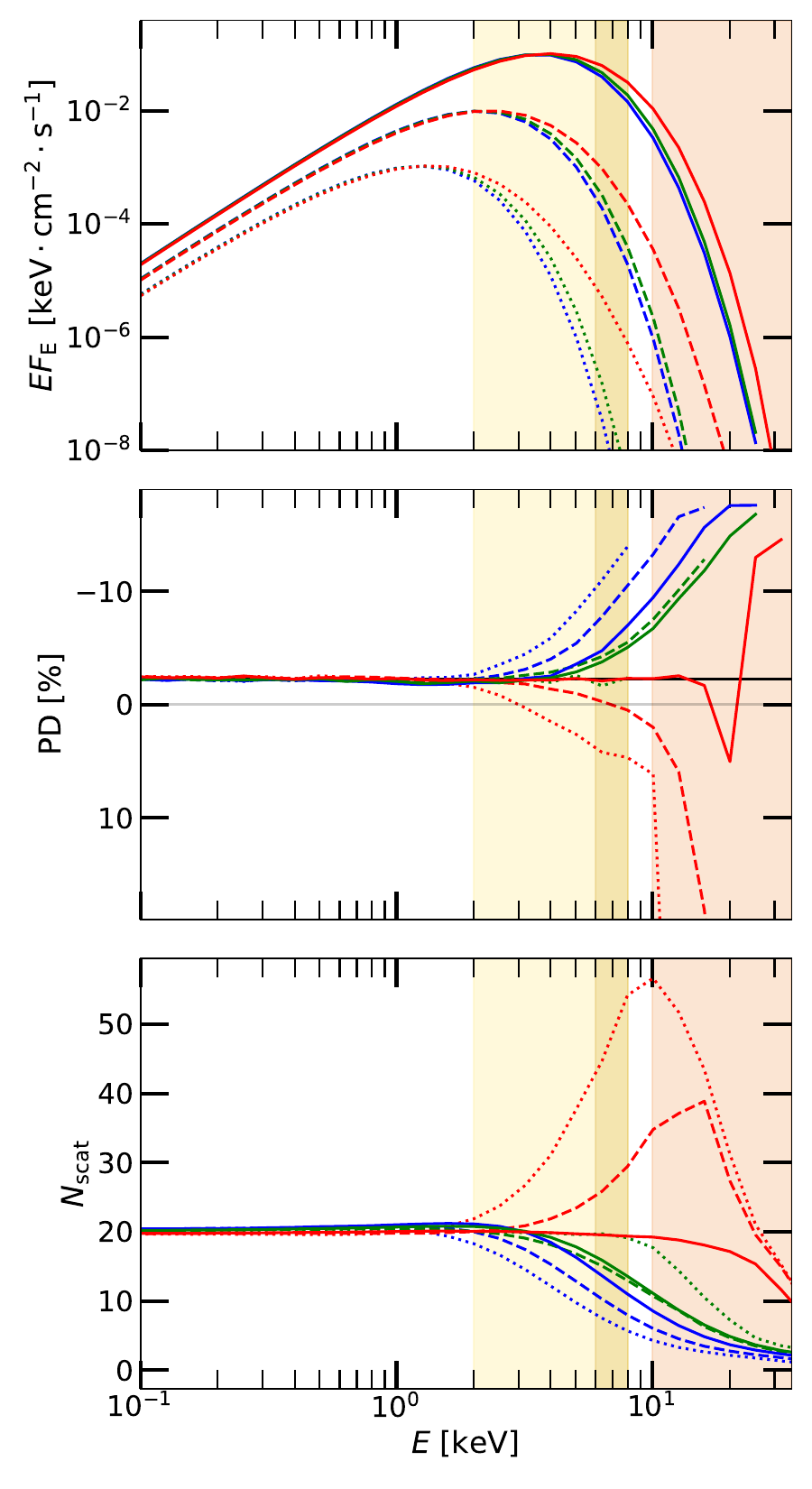}};
            \node[anchor=north west, xshift=-3mm, yshift=1mm, font=\bfseries] at (img.north west) {(a)};
        \end{tikzpicture}
    \end{subfigure}
    \hfill
    \begin{subfigure}[t]{0.575\textwidth}
        \centering
        \begin{tikzpicture}
            \node[inner sep=0] (img)
                {\includegraphics[width=\linewidth,trim=0 0 0 0,clip]{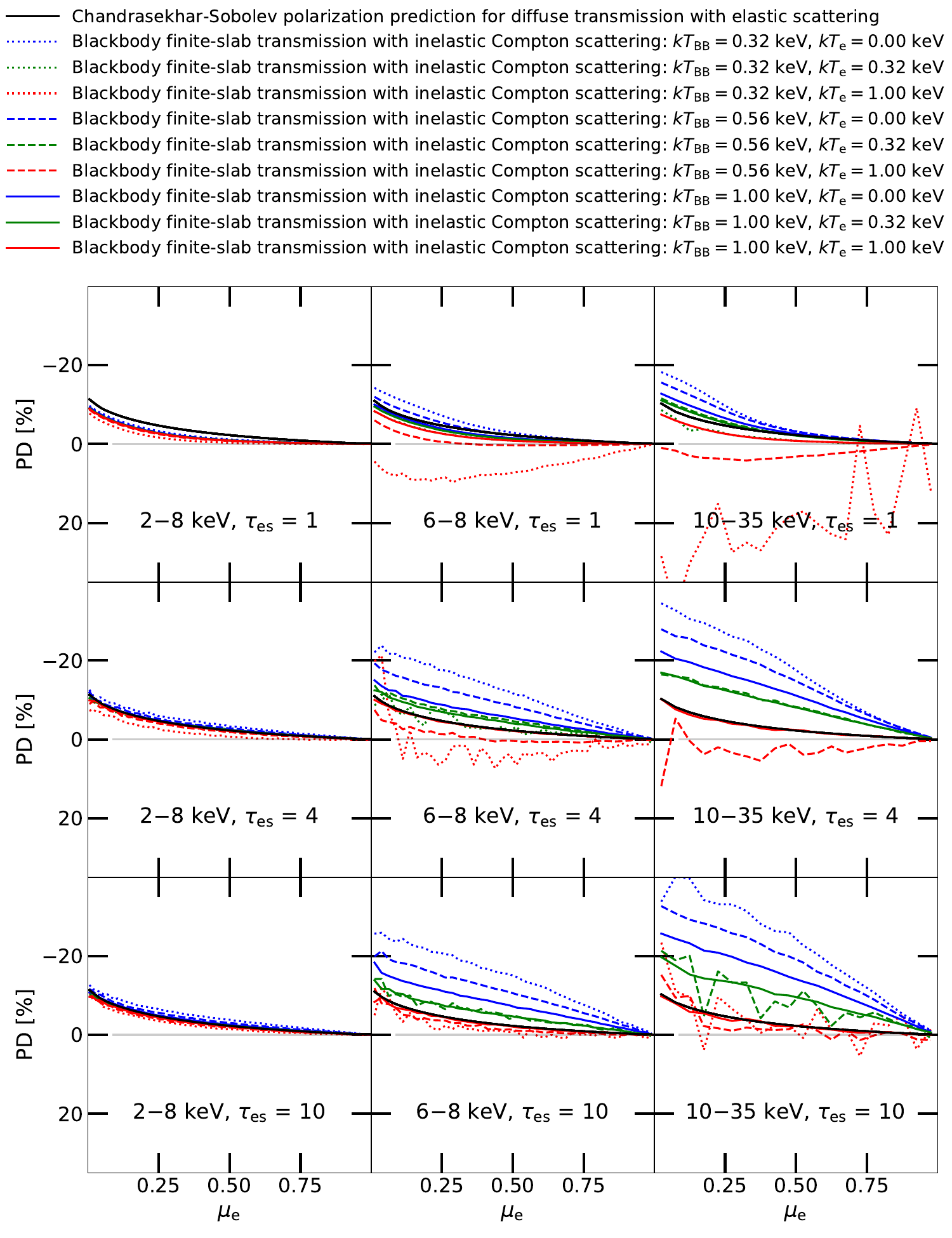}};
            \node[anchor=north west, xshift=-2.5mm, yshift=-29mm, font=\bfseries] at (img.north west) {(b)};
        \end{tikzpicture}
    \end{subfigure}
    \caption{Difference between the Chandrasekhar--Sobolev transmission polarization prediction for semi-infinite electron-scattering atmosphere (black solid lines) and the emission from a plane-parallel slab of finite $\tau_\mathrm{es}$ for inelastic Compton scattering, for different incident blackbody spectra (dotted lines: $kT_\mathrm{BB} = 0.32$ keV, dashed lines: $kT_\mathrm{BB} = 0.56$ keV, solid lines: $kT_\mathrm{BB} = 1$ keV) and temperatures of the slab (blue lines: $kT_\mathrm{e}=0$ keV, green lines: $kT_\mathrm{e}=0.32$ keV, red lines: $kT_\mathrm{e}=1$ keV), integrated in the azimuth. On the left (a), we show the spectra, polarization, and the obtained number of scatterings versus X-ray energy (from top to bottom) for $\mu_\mathrm{e} = 0.5$, $\tau_\mathrm{es} = 4$. On the right (b), we show the integrated polarization versus $\mu_\mathrm{e}$ in 2--8 keV, 6--8 keV, 10--35 keV (from left to right, indicated with shaded regions in (a) panels) for different $\tau_\mathrm{es} = 1, 4, 10$ (from top to bottom). Some curves in (b) are omitted due to low flux and high numerical noise.}
    \label{fig:transmission}
\end{figure*}

Figure~\ref{fig:transmission}a shows MC examples of single-temperature blackbody spectra, characterized by $kT_\mathrm{BB}$, transmitted through a cold and warm slab with lower or comparable $kT_\mathrm{e}$. The results are shown for a fixed $\tau_\mathrm{es} = 4$, and a selected cosine of inclination $\mu_\mathrm{e} = \cos{\delta_\mathrm{e}} = 0.5$. As expected for the spectra, due to the down- and up-scattering energy shifts upon each scattering event, the colder/hotter the electrons, the softer/harder the transmitted spectrum of the same incident temperature, respectively. In the soft X-rays (below $\sim 2$ keV for the lowest incident $T_\mathrm{BB}$ shown), the Compton energy shifts no longer play a role and the transmitted spectra converge to the incident spectra. For polarization in the soft X-rays, the PD follows the Chandrasekhar--Sobolev value for a semi-infinite atmosphere (displayed for the selected viewing angle in black in Fig.~\ref{fig:transmission}a), because the inelasticity of scattering plays a negligible role for the cases shown and the slab is already effectively optically thick \citep{SunyaevTitarchuk1985}. In the mid and hard X-rays, the PD departs from the Chandrasekhar--Sobolev values, unless the radiation is approximately at the temperature of the electrons in the slab. That is the case for the green dotted lines ($kT_\mathrm{BB} = 0.32$ keV, $kT_\mathrm{e} = 0.32$ keV) and red solid lines ($kT_\mathrm{BB} = 1$ keV, $kT_\mathrm{e} = 1$ keV) up to a frequency where the flux falls below detectable values ($\gtrsim5$ orders of magnitude below its peak X-ray value).

The rest of the curves show that towards hard X-rays, the inelasticity causes departures from the Chandrasekhar--Sobolev values. The departures are either in the direction of 
\begin{enumerate}
    \item[(1)] increasing a negative PD,
    \item[(2)] or decreasing a negative PD and then increasing a positive PD after a transition through $\mathrm{PD}=0$.
\end{enumerate}
The first branch of solutions is due to cold Comptonization: electron temperature is, on average, lower than the temperature of the incident radiation. The latter branch of solutions is due to warm Comptonization: electrons are, on average, warmer than the incident radiation. To show in detail why the two branches in PD occur, we also plot in Fig.~\ref{fig:transmission}a the average scattering order per observed energy bin, $N_\mathrm{scat}$, versus energy. In the cold branch, the photons get on average down-scattered by colder electrons. $N_\mathrm{scat}$ decreases with energy in the mid and hard X-rays, because the photons that originate in the high-energy tail are shifted by initially large, and then smaller $\Delta E$, according to (\ref{eq:compton_drift}). The bins at relatively higher energies with lower number of scatterings show higher PD due to the less uniform distribution of scattering planes and angles in such configurations. The cumulative Compton recoil is effective enough to cause a steep rise of PD with energy even shortly after the spectral peak in $EF_\mathrm{E}$ of the transmitted radiation. The lower the blackbody temperature, the softer the spectral peak and the lower the critical energy where the PD departs from the Chandrasekhar--Sobolev value. Within the characteristic range of departure points set by the incident spectral shape, the PD departs at lower energies in the cold branch when there is a larger difference between the electron and radiation temperature (compare the blue and green lines of the same line type in Fig.~\ref{fig:transmission}a).

The same Comptonization effect, but due to predominant up-scattering continues towards the warm branch (compare the red lines in Fig.~\ref{fig:transmission}a with the blue and green ones of the same line type). In the warm branch, higher electron temperatures relative to the radiation temperature cause a peak in $N_\mathrm{scat}$ with cumulated up-scatterings (originally, on average, lower energy photons) and down-scatterings (originally, on average, higher energy photons). The counter inelastic effect to down-scattering causes a depolarization at higher energies from the departure point from the Chandrasekhar--Sobolev value, and then the trend continues towards hard X-rays with an opposite sign due to the same mechanisms as multiple orders of up-scatterings impact X-ray polarization in hot coronae in the slab geometry, which result in polarization angles aligned with the slab normal \citep[e.g.,][]{Poutanen1996}. At low energies, $N_\mathrm{scat}$ converges to $\sim 20$, which is given by the selected thickness of the slab and the isotropy law for the external source. If we had chosen incident X-ray spectra of lower temperatures, the differences in the highlighted bands would have been larger and would start, analogously, at softer X-ray energies, but the flux in the mid X-ray band---a conditioner for polarization detectability---would also decrease in the high-energy spectral tail.

Figure~\ref{fig:transmission}b shows how the cold and warm branches in PD behave for different $\mu_\mathrm{e}$ and $\tau_\mathrm{es}$. Using the same MC approach, we integrate the results in different energy bands shown by the yellow and orange shaded regions in Fig.~\ref{fig:transmission}a that correspond to detector ranges of current or planned X-ray polarimeters \citep[e.g.,][]{Weisskopf2022, Soffitta2026}. In $\mu_\mathrm{e}$, the PD with included Compton scattering effects follow the trend of PD($\mu_\mathrm{e}$) set by the elastic Chandrasekhar--Sobolev profile (shown in black in Fig.~\ref{fig:transmission}b). Most clearly for the 2--8 keV band (left column) where the spectral shape weights higher the \textit{softer} photons in that band, which have the Chandrasekhar--Sobolev polarization for $\tau_\mathrm{es} \gtrsim 3$. This is clear from the comparison with the 6--8 keV band (middle column). In the highest energy band shown, 10--35 keV, the departures from Thomson-scattering predictions in both cold and warm branches are significant, particularly for low $\mu_\mathrm{e}$. Lower $\tau_\mathrm{es}$ causes the cumulative Compton recoil effect in the cold branch to be milder, because the photons do not accumulate as many past scattering events as shown for $\tau_\mathrm{es} = 4$ in Fig.~\ref{fig:transmission}a. For higher $\tau_\mathrm{es}$, the departure from the Chandrasekhar--Sobolev profile in the cold branch becomes prominent already in the 6--8 keV band. On the contrary, the warm Comptonization causes a more steep modification of PD with energy for low $\tau_\mathrm{es}$. And it shows saturation with high $\tau_\mathrm{es}\gtrsim 10$ towards the Chandrasekhar-Sobolev values in the X-ray band due to the accumulated high scattering orders effectively converging to the elastic scattering predictions, as described in slab corona studies where hot electrons up-scatter soft thermal seed photons \citep[e.g.,][]{SunyaevTitarchuk1985}.

\subsection{Reflection}\label{reflection}

\begin{figure*}[t]
    \centering
    \begin{subfigure}[t]{0.41\textwidth}
        \centering
        \begin{tikzpicture}
            \node[inner sep=0] (img)
                {\includegraphics[width=\linewidth,     
                   trim={0cm 0.35cm 0cm 0cm},
                   clip]{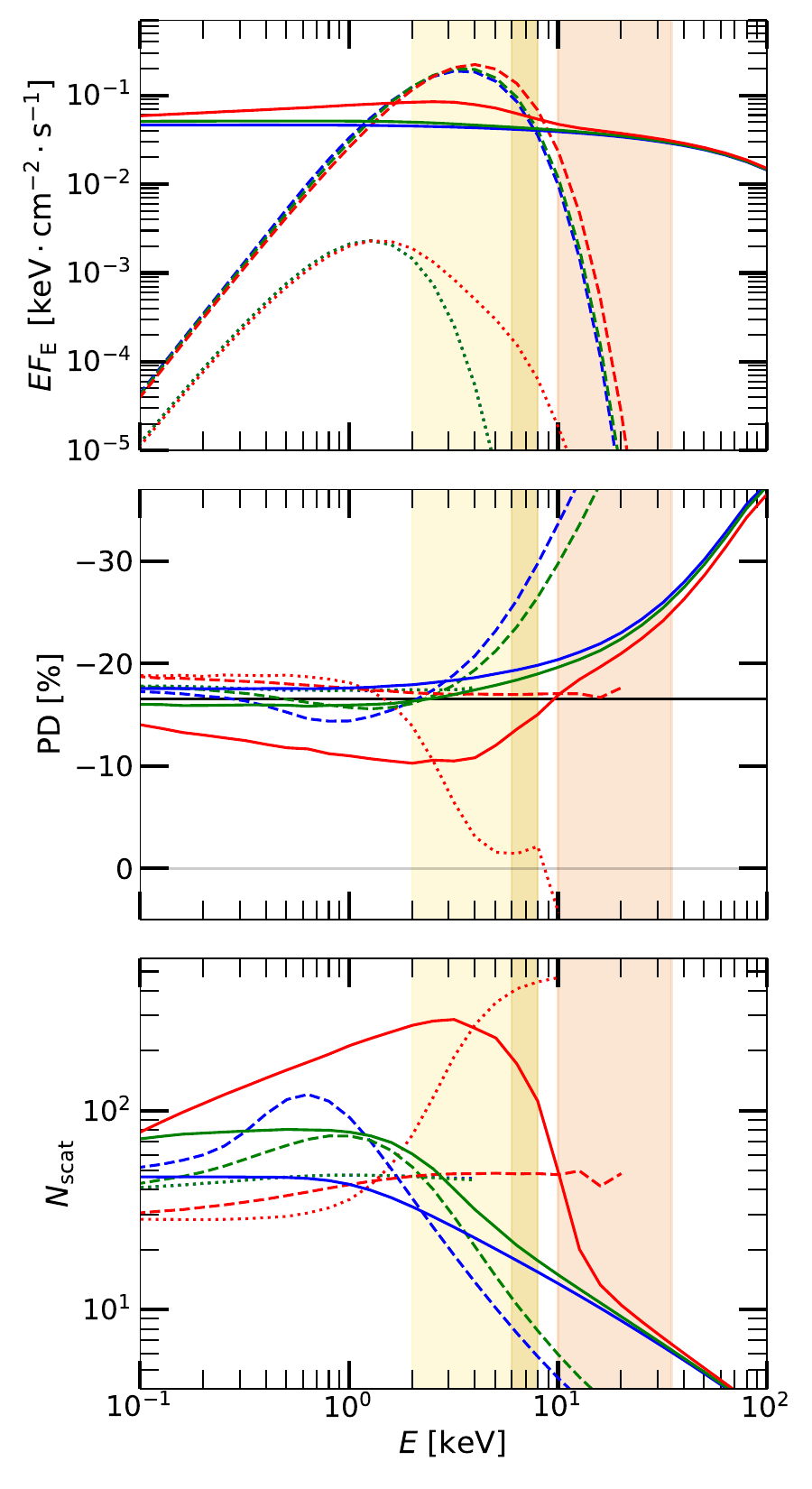}};
            \node[anchor=north west, xshift=-3mm, yshift=1mm, font=\bfseries] at (img.north west) {(a)};
        \end{tikzpicture}
    \end{subfigure}
    \hfill
    \begin{subfigure}[t]{0.575\textwidth}
        \centering
        \begin{tikzpicture}
            \node[inner sep=0] (img)
                {\includegraphics[width=\linewidth]{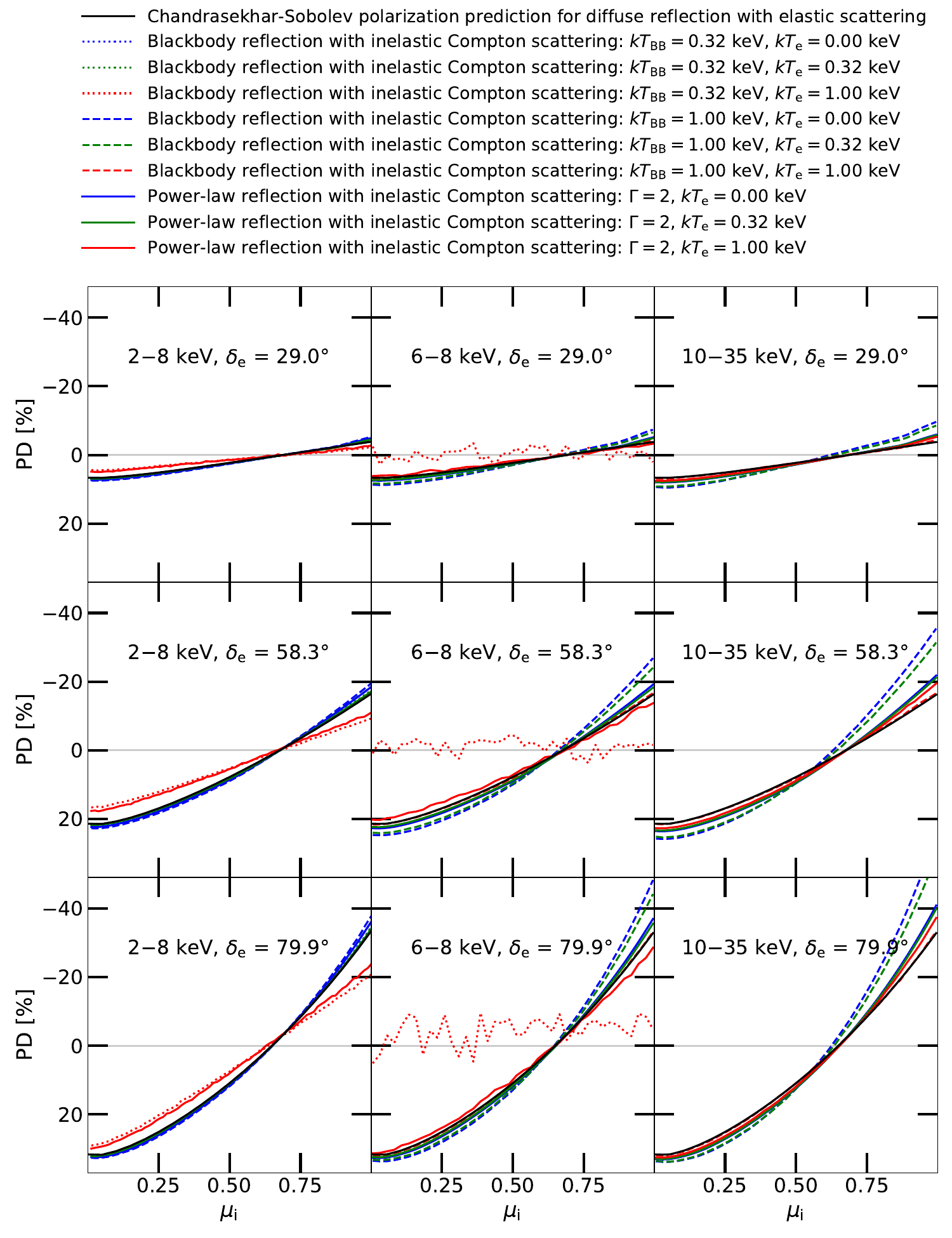}};
            \node[anchor=north west, xshift=-3mm, yshift=-28mm, font=\bfseries] at (img.north west) {(b)};
        \end{tikzpicture}
    \end{subfigure}
    \caption{Difference between the Chandrasekhar-Sobolev diffuse reflection polarization prediction (black solid lines) and calculations with Comptonization for different incident X-ray spectra (dotted lines: blackbody with $kT_\mathrm{BB} = 0.32$ keV, dashed lines: blackbody with $kT_\mathrm{BB} = 1$ keV, solid lines: power law with $\Gamma = 2$) and temperatures of the slab (blue lines: $kT_\mathrm{e}=0$ keV, green lines: $kT_\mathrm{e}=0.32$ keV, red lines: $kT_\mathrm{e}=1$ keV), integrated in the azimuth for plane-parallel electron-scattering optically thick slabs. On the left (a), we show the spectra, polarization, and the obtained number of scatterings versus X-ray energy (from top to bottom) for $\mu_\mathrm{e} = 0.525$, $\mu_\mathrm{i} = 1$. On the right (b), we show the integrated polarization versus $\mu_\mathrm{i}$ in 2--8 keV, 6--8 keV, 10--35 keV (from left to right, indicated with shaded regions in (a) panels), for different $\delta_\mathrm{e} = 29^\circ, 59^\circ, 80^\circ$ (from top to bottom). Some curves in (b) are omitted due to low flux and high numerical noise.}
    \label{fig:reflection}
\end{figure*}

Figure~\ref{fig:reflection} shows equivalent quantities to Fig.~\ref{fig:transmission}, but for reflection geometry. We compare with MC simulations the Chandrasekhar--Sobolev predictions for diffuse reflection from semi-infinite electron-scattering atmospheres. We chose $\tau_\mathrm{es} = 20$ in the simulations to represent an optically thick reflection. We tested that higher $\tau_\mathrm{es}$, which significantly slows down the MC simulations, still affects the average scattering order obtained and the very high-energy tail in spectra and PD, where the inelastic scattering causes the strongest departures from the Chandrasekhar--Sobolev predictions. Figure~\ref{fig:reflection}a shows the MC results specifically for the cosine of incident inclination $\mu_\mathrm{i} = \cos{\delta_\mathrm{i}} = 1$ and $\mu_\mathrm{e} = 0.525$. We show two incident blackbody spectra with different $kT_\mathrm{BB}$ (selected from Fig.~\ref{fig:transmission}) and one incident power-law spectrum with a photon index $\Gamma = 2$, versus the cold and warm electron temperatures as in Fig.~\ref{fig:transmission}.

The same inelastic scattering mechanism causes departures from the classical results, which is particularly prominent at high energies where for all cases shown (apart from dotted lines that are cut due to lack of high-energy photons) the photons on average loose energy (cold Comptonization) and the (negative) PD dramatically increases with energy. For a reflecting power-law, the high-energy tail in PD is shifted to higher energies than for the blackbody with $kT_\mathrm{BB}=1$ keV, because its spectra are less steeply declining at higher energies and the high-energy photon reservoir causes higher $N_\mathrm{scat}$ values there. Hence, the effect is dependent on the incident spectral energy distribution. The red solid curves ($\Gamma = 2$, $kT_\mathrm{e} = 1$ keV) show a peak in spectra, a corresponding peak in $N_\mathrm{scat}(E)$ and a dip in PD($E$), as the reflecting warm slab predominantly up-scatters the impinging photons below the peak, and predominantly down-scatters the impinging photons above the peak. Thus, in such case, the reflected PD from soft to hard X-rays first decreases from the elastic scattering prediction (up-scattering), meets it near 10 keV, and exceeds it above (down-scattering). The same effect, to a smaller extent, can be seen from the blue and green dashed curves (1 keV blackbody reprocessed by a colder slab) and in the low-energy (up-scattered) part on the dotted red curve (0.32 keV blackbody reprocessed by a warm slab) cut at higher energies due to lack of photons. Similarly to transmission, the cases with equivalent electron and radiation temperatures (green dotted and red dashed lines) follow the classical diffuse reflection value closely in the X-rays due to the energy balance mechanism explained in Section \ref{compton}. The blue dotted curve is almost identical to the green dotted curve, because for $kT_\mathrm{BB} = 0.32$ keV, the inelastic effects for $kT_\mathrm{e} \lesssim 0.32$ keV are ineffective in the studied energy range and the polarization follows the elastic predictions.

Figure~\ref{fig:reflection}b shows the PD, integrated in the same energy ranges as in Fig.~\ref{fig:transmission}b versus $\mu_\mathrm{i}$ and $\mu_\mathrm{e}$. The angles define the prevailing scattering geometry, which is imprinted in the classical elastic reflection predictions (black solid curves), to which (apart from the strong warm Comptonization examples in red dotted and red solid lines) the cases for inelastic scattering shown adhere when integrated in the 2--8 keV band---again largely due to higher soft photon weighting. The largest departures from classical diffuse reflection occur at high $\mu_\mathrm{i}$, because, if the photons are injected perpendicularly to the slab surface, they enter forward-scattered deeply, experience on average higher scattering orders than for lower $\mu_\mathrm{i}$, and the cumulative Compton recoil is more effective. The departures from the black curves appear again, to a smaller extent, for very low $\mu_\mathrm{i}$ where only a few scattering events on average occur before the photons escape the slab \citep{Podgorny2026}, and the results imprint the differences between elastic and inelastic individual scattering events (both geometrical and energetic redistributions).

Reflection from cold matter naturally involves very few scattering events due to absorption, and is often approximated by the classical elastic single-scattering predictions. It was already shown in \cite{Matt1993} and \cite{Poutanen1996b}, including the energy-dependent photo-electric absorption and spectral-line effects, that for the reflecting power-law spectrum, the cold Compton recoil can cause strong deviations in PD from the elastic predictions at already $\gtrsim 3$ keV. The comparison with Chandrasekhar's values is shown directly in Fig.~A.3 of \cite{Podgorny2026}, which reproduces the results from \cite{Poutanen1996b} with MC and includes a discussion on the inapplicability of the classical elastic single-scattering (and diffuse) reflection prescriptions in the X-rays.

\section{Summary and discussion}\label{conclusions}

We demonstrated to what extent the inelastic scattering in cold and warm atmospheres causes departures from classical Chandrasekhar--Sobolev polarization predictions in the X-rays. Assuming a homogeneous slab filled with electrons only, significant departures from the classical Thomson-scattering predictions (a few to even tens of \% in PD, and in some cases changing its sign) appear in the hard X-rays (above 10 keV) and in several cases in the mid X-ray band (1--10 keV) already. In transmission, these occur for soft primary spectra, high $\tau_\mathrm{es}$ for electrons colder than the radiation or small $\tau_\mathrm{es}$ for electrons hotter than the radiation, and generally high relative difference between the radiation and electron temperatures in the reprocessing medium. In optically thick reflection, these occur for soft primary spectra, high $\mu_\mathrm{i}$, and high relative difference between radiation and electron temperatures.

Accretion-disk atmospheres of accreting compact objects, even if assumed fully ionized, exhibit complex electron temperature, electron density, free-free emissivity, and free-free opacity vertical profiles, which also vary with the radial distance from the disk center and, more generally, depend on the global accretion conditions \citep[e.g.,][]{HubenyEtAl2001, Blaes2006, Kyriazis2026}. Therefore, our selected passive homogeneous slab examples are only indicative of the basic expected inelastic scattering effects on the emergent rest-frame mid- and hard-X-ray polarization. The depth of formation $\tau_\nu^\ast$, which we have shown to be a strong X-ray polarization driver even for optically thick transmitting atmospheres, and the related color-correction factor and black-hole spin estimates \citep{Davis2006}, are dependent, e.g., on the detailed role of energy dissipation, magnetic reconnection, and turbulence inside the atmosphere \citep{DavisEtAl2005, Davis2009}. Thus, self-consistently calculated atmospheres with inelastic scattering in global accretion models, if compared with X-ray polarimetric data, could indirectly constrain the role of magnetic fields beyond existing limits from the lack of observed Faraday rotation in X-rays \citep{Barnier2024, Krawczynski2026}. The warm Comptonization effects on X-ray polarization could in principle test the existence of a warm coronal skin on the disk atmosphere; however, the effects of inelasticity seen in polarization are small for optically thick warm layers, which is the typical prediction ($\tau_\mathrm{es}\sim 10$) for warm coronae \citep{Gronkiewicz2020}. Furthermore, inelastic scattering effects in rest-frame reflection can play a non-negligible role in X-ray polarimetric studies of accretion-disk self-irradiation \citep{Podgorny2026}, which can additionally be used to constrain the disk properties and the spin \citep{Schnittman2009, West2023, Steiner2024}. Thus, by accounting for the outlined cold- and warm-Comptonization effects in accretion-disk atmospheres, we expect energy-resolved X-ray polarimetry to indirectly constrain the physical conditions of accretion and potentially reassess black-hole X-ray binary spin estimates obtained either from purely spectral continuum fitting in the soft state \citep[e.g.,][]{McClintock2014} or from existing spectro-polarimetric methods based on models using the classical elastic-scattering predictions (see Section \ref{introduction}). In the latter case, incorporating cold and warm Comptonization effects will eliminate some systematic errors and provide a more robust diagnostic.

Several \ixpe\ observations of accreting black holes and weakly-magnetized neutron stars showed a rise of PD with X-ray energy in the 2--8 keV range together with a constant PA with X-ray energy \citep{Dovciak2024, Ursini2024}, which can be in part caused by the rest-frame inelastic scattering, in addition to the more commonly accepted origin via mixing of differently polarized spectral components. Some magnetized accretion disk solutions, resulting in local temperature decrease with height, large color-correction factors [\citealp[$f_\mathrm{col}\simeq T_\mathrm{BB}/T_\mathrm{e}\propto (y_\mathrm{eff}/T_\mathrm{e})^{1/9}$,][]{Davis2006}], and $\tau_\nu^\ast\gg 10$ at the still highly-ionized inner disk radii \citep{Begelman2007, Davis2009} may lead to an enhanced role of the cold Comptonization branch in transmission. This raises the possibility for explanation of the energy dependence and remarkably high value of the soft-state 2--8 keV polarization in 4U1630$-$47 \citep[][where the authors showed that the only plausible fits of polarization, assuming a Novikov-Thorne disk, required relatively high $\tau_\mathrm{es}$, high ionization, and a cold outflowing atmosphere]{Ratheesh2024} and the subsequent polarization dilution in the \textit{warmer}, steep-power-law state \citep{RodriguezCavero2023}. However, to test this hypothesis, a set of sophisticated non-LTE rest-frame atmospheric polarized radiative transfer simulations, taking into account the hydrostatic equilibrium and realistic energy dissipation, coupled to relativistic radiative-transfer codes, as pioneered by \cite{Davis2009}, and possibly self-consistently calculated self-irradiation effects (also affecting the atmospheric structure), need to be built. Moreover, the photo-electric absorption and spectral-line effects, inevitably occurring towards the outer disk, may impact the observed polarization \citep{Taverna2021, Marra2025}. If so, it will further complicate the building of future X-ray spectro-polarimetric models and the revelation of the full diagnostic power of X-ray polarimetry. 

Future X-ray polarimeters, such as eXTP \citep{extp} and EXPO \citep{Soffitta2026}, will further tighten the possible constraints arising from modeling structured atmospheres with inelastic scattering. In particular, EXPO is planned to operate in the hard 10--35 keV band, where we predict the strongest polarization departures from classical expectations. But, at least for soft-state X-ray binary systems, the source flux in this band is orders of magnitude lower relative to the peak flux, which reduces the probability of detection of polarization in the hard band.

\begin{acknowledgements}

J.Pod., M.D., and M.G. were supported by GACR project 26-22614S and acknowledge institutional support from RVO:67985815. J.Pod. acknowledges VSB Technical University of Ostrava, IT4Innovations National Supercomputing Center, Czech Republic, for awarding this project access to the LUMI supercomputer, owned by the EuroHPC Joint Undertaking, hosted by CSC (Finland) and the LUMI consortium through the Ministry of Education, Youth and Sports of the Czech Republic through the e-INFRA CZ (grant ID: 90254). J.Pou. and A.V. were supported by the Research Council of Finland grants 355672 and 372881 and Centre of Excellence in Neutron-Star Physics (grant 374064). Nordita is supported in part by NordForsk. This research was supported by the International Space Science Institute (ISSI) in Bern, through ISSI International Team project `Maximizing the Potential of X-ray Polarimetric Data to Understand Accreting BHs' (ISSI Team project \#25-649).

\end{acknowledgements}

\bibliography{refs}
\bibliographystyle{yahapj}

\end{document}